Superconductivity-induced Resonance Raman Scattering

in Multi-layer High-$T_c$ Superconductors

Mikhail Limonov,* Sergey Lee, and Setsuko Tajima,

[1]*SRL-ISTEC, 10-13, Shinonome 1-Chome, Koto-ku, Tokyo 135-0062, Japan*

Akio Yamanaka

[2]*Chitose Institute of Science and Technology, Chitose, Hokkaido 066-8655, Japan*

Resonant Raman scattering below $T_c$ has been discovered in several Bi-, Hg-, Tl-based high-$T_c$ superconductors with three or four $CuO_2$-layers. For $Bi_2Si_2Ca_2Cu_3O_{10+\delta}$, we found an unexpected crossover of the pair-breaking peak in the $A_{1g}$-spectrum from a broad bump at $\hbar\omega = 6k_BT_c$ for $E_{exc} = 2.54 eV$ to a sharp peak at $\hbar\omega = 8k_BT_c$ for $E_{exc} = 2.18 eV$, together with a strong enhancement of the Ca-phonons. Under resonant conditions, the relative positions of the pair breaking peaks in $A_{1g}$, $B_{1g}$, and $B_{2g}$ channels are $2\Delta(A_{1g}) = 2\Delta(B_{1g}) > 2\Delta(B_{2g})$. This relation implies that the $A_{1g}$ Raman channel is free from the Coulomb screening effect, just as predicted theoretically for a d-wave multi-layer superconductor but have never been observed experimentally thus far. The observed resonance effect is the evidence that the electronic state in the inner $CuO_2$-planes is different from that of the outer $CuO_2$-planes.

Correspondent author : S.Tajima (tajima@istec.or.jp)

**I. INTRODUCTION**

A layered structure together with a strong electron correlation in high-$T_c$ superconducting cuprates (HTSC) produces a large anisotropy in their electronic state, resulting in an incoherent charge transport in the out-of-plane direction [1] which cannot be expected from conventional band theory. In the superconducting state, owing to a short coherence length, the system can be regarded as a stack of Josephson-coupled-layers where each $CuO_2$-plane is an independent superconducting layer [2]. In the case of compounds with multiple $CuO_2$-layers in a formula unit, the multi-sheet effect manifests itself in the peculiar charge response such as the double Josephson plasma accompanied with an additional absorption peak [3]. Another report of multi-sheet effect is the splitting of the Fermi surface observed in the angle-resolved photoemission spectroscopy on the heavily overdoped $(Bi,Pb)_2Sr_2CaCu_2O_8$ [4]. In the nuclear magnetic resonance experiment, it is indicated that the doping level of the inner $CuO_2$-planes is lower than that of the outer planes [5]. So far, most of the theories for HTSC deal with the electronic state only within the $CuO_2$-planes, ignoring the structure in the out-of-plane direction. However, it is empirically known that layer-stacking strongly affects some of the electronic properties. For example, a simple but crucially important question is why $T_c$ correlates with the number of $CuO_2$-planes in a formula unit. Therefore, theories dealing only with the two-dimensional $CuO_2$-planes are insufficient for describing the essential features of the whole HTSC system.

The effect of the multi-layer coupling on Raman scattering spectra has been widely discussed in relation to the polarization dependence of the superconducting gap feature, although the initial discussion was focused on the theory based on a single sheet Fermi surface (FS). The electronic Raman efficiency in the superconducting state was derived by Klein and Dierker [6]. For a single sheet Fermi surface (FS), in the $q \rightarrow 0$ limit the efficiency S is given by

$$S \propto \left[ 1 - e^{-\hbar w / k_B T} \right] \mathrm{Im}\left[ \left\langle \boldsymbol{g}(\mathbf{k})^2 \, \boldsymbol{l}(\mathbf{k}) \right\rangle_F - \frac{\left\langle \boldsymbol{g}(\mathbf{k}) \boldsymbol{l}(\mathbf{k}) \right\rangle_F^{\,2}}{\left\langle \boldsymbol{l}(\mathbf{k}) \right\rangle_F} \right], \qquad (1)$$

where $\langle \, \rangle_F$ denotes an average over the FS and $\lambda(\mathbf{k})$ is the complex tsuneto function. For a nearly cylindrical FS the imaginary part of $\lambda(\mathbf{k})$ at T=0 has a form

$$\mathrm{Im}\left[ \boldsymbol{l}(\mathbf{k}) \right] = \frac{4\left|\Delta(\mathbf{k})\right|^2}{\sqrt{w^2 - 4\left|\Delta(\mathbf{k})\right|^2}} \ . \qquad (2)$$

The Raman vertex $\gamma$ is described as

$$\boldsymbol{g}(\mathbf{k}) = \mathbf{e}^L \cdot \mathbf{e}^S + \frac{1}{m}\sum_b \left[ \frac{\langle \mathbf{k} | \mathbf{p} \cdot \mathbf{e}^S | b, \mathbf{k} \rangle \langle b, \mathbf{k} | \mathbf{p} \cdot \mathbf{e}^L | \mathbf{k} \rangle}{\boldsymbol{e}(\mathbf{k}) - \boldsymbol{e}_b(\mathbf{k}) + \hbar w_L} + \frac{\langle \mathbf{k} | \mathbf{p} \cdot \mathbf{e}^L | b, \mathbf{k} \rangle \langle b, \mathbf{k} | \mathbf{p} \cdot \mathbf{e}^S | \mathbf{k} \rangle}{\boldsymbol{e}(\mathbf{k}) - \boldsymbol{e}_b(\mathbf{k}) - \hbar w_S} \right], \qquad (3)$$

where $\mathbf{e}^L$ and $\mathbf{e}^S$ denote the polarization vectors of the incident and scattering light, respectively. When the incident photon energy $\hbar\omega_L$ is much smaller than the energy difference $\varepsilon(\mathbf{k}) - \varepsilon_b(\mathbf{k})$ between the initial state $\mathbf{k}\rangle$ and intermediate state $|b,\mathbf{k}\rangle$, the Raman vertex is reduced to an inverse effective mass tensor $\mu^{-1}(\mathbf{k})$ of the relevant band $\varepsilon(\mathbf{k})$,

$$\boldsymbol{g}(\mathbf{k}) = \boldsymbol{m}^{-1}(\mathbf{k}) = \frac{m}{\hbar^2}\sum_{\boldsymbol{a},\boldsymbol{b}} e_{\boldsymbol{a}}^L \frac{\P^2 \boldsymbol{e}_{\mathbf{k}}}{\P k_{\boldsymbol{a}}\,\P k_{\boldsymbol{b}}} e_{\boldsymbol{b}}^S \ , \qquad (4)$$

where we take $\alpha$, $\beta = x, y$ for two-dimensional system.

The second term in Eq.(1) gives rise to a correction of the long range Coulomb effect, so-called screening effect. For structures with $D_{4h}$ symmetry, there are three Raman active channels regarding to the basic $ab$-plane with $A_{1g}$, $B_{1g}$ and $B_{2g}$ symmetries. The screening term $\langle \gamma\lambda \rangle_F$ vanishes for the $B_{1g}$ and $B_{2g}$ components but not for $A_{1g}$. This symmetry dependence of the screening effect on the Raman was experimentally observed in many HTSC. For example, in $Bi_2Si_2CaCu_2O_{8+\delta}$ (Bi-2212), peaks were found at about 520cm$^{-1}$ in the $B_{1g}$, 470cm$^{-1}$ in the $B_{2g}$, and 370cm$^{-1}$ in the $A_{1g}$ spectra, that is, the peak energy $\Delta(B_{1g}) > \Delta(B_{2g}) > \Delta(A_{1g})$ [7]. The symmetry dependence of the peak energy as well as the spectral profile is well explained in terms of a d-wave gap $\Delta(k) = \Delta_0 \cos 2\varphi$ [7,8], for which the maximum in the pair-breaking peak is predicted to appear sharply at $\omega = 2\Delta_0$ for the $B_{1g}$ channel, while a rather board peak is located at a lower frequency $\omega = 1.3\Delta_0$ for the $B_{2g}$ channel, reflecting the states near the gap node. The screening effect significantly smears the $A_{1g}$ peak and leads to the low frequency shift of the peak.

Krantz and Cardona have pointed out that this screening effect scenario is valid only in single $CuO_2$-layer HTSC, or in the case of completely degenerated bands for the layers [9]. If the bands are non-degenerated, the $A_{1g}$ channel includes the interband Raman process, in addition to the intraband one. The interband scattering, which originates from a mass difference $\mu_i(\mathbf{k})^{-1} - \mu_j(\mathbf{k})^{-1}$ in different sheets of the FS [10], should lead to intense unscreened $A_{1g}$ scattering with a peak $\omega = 2\Delta_0$, as in the $B_{1g}$-channel: $\Delta \quad (B_{1g}) = \Delta(A_{1g})$.

Devereaux and co-workers have theoretically examined this problem in detail for a bilayer system [11, 12] and revealed that the screening effect is still important and smears an intense $A_{1g}$ peak at $\omega = 2\Delta_0$, when the interlayer coupling is not very strong. Simultaneously, it has been recognized that the line shape of the $A_{1g}$ scattering is sensitive either to the interlayer coupling or to the FS structure. However, double layer HTSC such as $YBa_2Cu_3O_{7-\delta}$ (Y-123) and Bi-2212 exhibit nearly the same $A_{1g}$ spectrum. For this problem, there have been many proposals [8,9, 11-17] including the complicated gap function [16] different from a simple d-wave and a correction by spin fluctuation [17]. But, it is still a puzzle why there is no clear effect of interlayer coupling in bilayer HTSC [18, 19]. In other words, accounting for the $A_{1g}$ Raman scattering features is crucial to understand the multi-layer effect on HTSC.

In this context, it is of great interest to investigate the $A_{1g}$ Raman scattering under resonant condition, because the interband Raman process is expected to be enhanced through the resonant part of the Raman vertex between different sheets of FS;

$$g_{ij}(\mathbf{k}) \approx \frac{1}{m} \frac{\langle i, \mathbf{k} | \mathbf{p} \cdot \mathbf{e} | b, \mathbf{k} \rangle \langle b, \mathbf{k} | \mathbf{p} \cdot \mathbf{e} | j, \mathbf{k} \rangle}{\left[ e_b(\mathbf{k}) - e_j(\mathbf{k}) \right]^2 - \left( \hbar w_L \right)^2}, \qquad (5)$$

where $\omega_L \approx \omega_S$ and $\mathbf{e}^L = \mathbf{e}^S \equiv \mathbf{e}$. The electronic Raman scattering has been intensively studied over a wide range of excitation energies in HTSC [20-25]. The resonant phenomena occur at around 2.0 eV and above 2.5 eV. To the best of our knowledge, however, no attempt has been devoted to investigate the $A_{1g}$ problem mentioned above. Recently, the superconductivity-induced changes of the phonon resonance have been found in a bilayer HTSC, suggesting that superconducting order parameter affects the optical transition at 2.1 eV [26].

In this paper, we report resonant Raman scattering of multi-layer HTSC. A strong resonant enhancement of the electronic Raman scattering appears near 2 eV and develops below $T_c$ only in HTSC with three or four $CuO_2$ layers. A strong $A_{1g}$ peak is observed at an energy $2\Delta(A_{1g})$ close to $2\Delta(B_{1g})$, which suggests unscreened scattering under the resonance condition. This is an indication of interband Raman scattering that is allowed for non-degenerated bands corresponding to the inner and outer $CuO_2$-layers. The contribution of the Ca-O vibration phonons to the interlayer coupling is also revealed.

## II. EXPERIMENT

In contrast to the intensive studies on Y-123 and Bi-2212, there have been only a few works on three or four layer HTSC, probably because of a lack of high quality single crystals. To extend the research of these materials, we prepared various samples of HTSC with multiple $CuO_2$-planes.

Single crystals of Bi-2212 with $T_c$=85K and $Bi_2Si_2Ca_2Cu_3O_{10+\delta}$ (Bi-2223) with $T_c = 109$ K were grown by a KCl flux technique [27]. A new crystal growth method for Bi-2223 was developed by optimizing the crucible material, phase and chemical composition of the precursor powder, heat treatment procedure, and evaporation rate of KCl. As a result, we obtained high quality Pb-free Bi-2223 single crystals with a size of 500x500x2 $\mu m^3$. The details of crystal growth are published elsewhere [28]. The Bi-2212 single crystal is in the slightly overdoped regime, while the doping level of the Bi-2223 crystal is close to the optimum.

A single-phase polycrystalline sample of $Tl_{0.5}Pb_{0.5}Sr_2Ca_2Cu_3O_{9+\delta}$ (Tl-1223) with $T_c=118$ K, the third member (n=3) of $TlSr_2Ca_{n-1}Cu_nO_{2n+3+\delta}$ homologous series [29], was prepared by "space-filling" modification of the encapsulation technique developed previously for the synthesis of Hg-based superconducting cuprates [30]. As-grown samples used for Raman measurements are in the overdoped regime. Polycrystalline samples of $HgBa_2Ca_2Cu_3O_{8+\delta}$ (Hg-1223), $Hg(Sr,Ba)_2Ca_3Cu_4O_{10+\delta}$ (Hg-1234) phases - the third (n=3) and fourth (n=4) members of Hg-based homologous series with general formula $Hg(Ba, Sr)_2Ca_{n-1}Cu_nO_{2n+2+\delta}$ [31], were obtained by a high-pressure technique (t=1000°C, P=5GPa). Highly homogeneous oxide precursor powders were prepared by spray-drying and thermal decomposition of the nitrate solutions, from which we obtained Hg-1223 ($T_c=133$ K) and Hg-1234 ($T_c=115$ K) polycrystals with purity higher than 90%. The carrier doping level of the former sample is nearly optimal, while the latter is a little underdoped. In all these multi-layer HTSC, it is generally difficult to change the doping range widely.

The Raman spectra were measured in the pseudo-back scattering configuration with a T64000 Jobin-Yvon spectrometer. The typical spectral resolution was 3cm$^{-1}$. For excitation several Ar$^+$-Kr$^+$ laser lines ranging from 1.9eV to 2.7eV were used. The power density was about 5 W/cm$^2$ on the sample surface, and the overheating was estimated to be less than 10K. All the measured spectra were corrected for the spectrometer sensitivity by comparison with that of $BaF_2$ and the contribution of the Bose factor has been removed. Hereafter we refer to a tetragonal $D_{4h}$ point group with X and Y taken parallel to the Cu-O bonds. In contrast to $B_{1g}(XY)$ and $B_{2g}(X'Y')$ Raman spectra, the $A_{1g}$ component cannot be exclusively measured by in-plane scattering geometries and was estimated as $I(A_{1g}) = I(XX) - I(X'Y')$.

## III. RESULTS AND DISCUSSION

Figures 1a and 1c illustrate the $A_{1g}$-Raman scattering for Bi2212 and Bi2223, respectively, using a blue laser line with an excitation energy $E_{exc}$=2.54eV. The frequencies for the three major phonons in the spectrum of Bi2223 are almost the same as those of Bi2212. This is reasonable because the crystal structures of these two compounds are nearly identical except for the number of $CuO_2$-planes. The electronic Raman scattering of Bi2223 is also similar to that of Bi2212. The low frequency part of the continuum exhibits a clear redistribution below $T_c$, owing to the formation of a superconducting gap. The gap peak is very broad and centered at 450 $cm^{-1}$ that is slightly larger than 390 $cm^{-1}$ for Bi2212, probably because of the higher $T_c$ of Bi2223.

At the orange excitation ($E_{exc}$=2.18eV), there appears a remarkable difference in the superconducting response between the two spectra. As shown in Figs.1b and 1d, in contrast to the almost unchanged spectrum of Bi2212, the 2Δ-peak in the Bi2223 spectrum is enhanced and located at a higher frequency than that for the blue excitation. Furthermore, very strong additional phonon lines are observed at 260 $cm^{-1}$ and 390 $cm^{-1}$.

In Fig.2a, the spectra at four different temperatures demonstrate a gradual development of the spectrum below $T_c$ peaking at 580 $cm^{-1}$. Figure 2b illustrates the temperature dependence of the Raman intensity at 580$cm^{-1}$ that is normalized at 1000$cm^{-1}$. It is clearly seen that a dramatic increase in the peak intensity as well as its resonance behavior sets in just below $T_c$.

Figure 3a displays the $A_{1g}$ spectra of Bi2223 at 10K for several excitation energies. The electronic intensity between 200 to 900 $cm^{-1}$ rapidly grows with decreasing $E_{exc}$, and consequently shows a strong 2Δ-peak at 580 $cm^{-1}$. Simultaneously, the weak phonon bumps in the blue spectrum become stronger and nearly symmetric lines at 260 $cm^{-1}$ and 390 $cm^{-1}$. The manifestation of these phonons is closely related to the growth of the 580 $cm^{-1}$ gap peak.

Comparing the $A_{1g}$-spectrum with the $B_{1g}$- and $B_{2g}$-spectra at 10K for $E_{exc}$=2.18eV (Fig.3b), we immediately find that the $A_{1g}$-peak position coincides with that for $B_{1g}$. Therefore, "in resonance", the relative positions of the pair breaking peaks in the $A_{1g}$, $B_{1g}$, and $B_{2g}$ channels is $2\Delta(A_{1g}) = 2\Delta(B_{1g}) > 2\Delta(B_{2g})$, as predicted theoretically for a d-wave multi-layer superconductor [9, 11, 12] but not observed to date. This implies that the resonant part of the $A_{1g}$ scattering is primarily free from screening effect due to the long range Coulomb force [12] and/or collective spin fluctuation [17]. Note that the profile and the peak position do not change with changing $E_{exc}$, in the $B_{1g}$- and $B_{2g}$-spectra (not shown here). The appearance of a high frequency bump at around $\omega = 3\Delta_0$ in the $B_{1g}$-polarization might originate from electronic excitations around the van Hove singularity [12, 32].

Figure 4a shows the resonance behavior of the height of the $2\Delta(A_{1g})$-peak at 580$cm^{-1}$ in Bi-2223, in comparison with the almost $E_{exc}$-independent behavior of the $2\Delta(A_{1g})$-peak at 390$cm^{-1}$ in Bi-2212. Here, the 2$\Delta$-peak intensities are normalized by the continuum intensity at 1000$cm^{-1}$. In Fig.4b, the renormalized phonon intensities $I_p = A\Gamma q^2$ [33, 34] for the 260 and 390 $cm^{-1}$ phonon lines in Bi-2223 are plotted as a function of $E_{exc}$. For Bi-2223, the resonant phenomenon is evident, although the data does not reach a maximum. The relevant energy level is presumably located at about 2eV or below. In contrast, the relative intensity of the $2\Delta(A_{1g})$-peak in Bi-2212 is almost independent of the excitation energy, as was observed in Tl-2201 [22].

The Raman vertex $\gamma$ has a resonance character if the photon energy $\hbar w_L$ (=$E_{exc}$) is close to the electronic excitation energy $\varepsilon_b$-$\varepsilon_j$, as is seen in eq.(5). This strongly enhances the Raman scattering efficiency S in Eq.(1) [10]. In Fig.4, the phonon intensity exhibits a similar $E_{exc}$-dependence. Therefore, it is supposed that the phononic Raman scattering is resonant to the same energy level.

The enhanced phonon lines at 260 and 390 $cm^{-1}$ in Bi-2223 can be ascribed to the $A_{1g}$-vibration of Ca with oxygen in the outer $CuO_2$-planes, following the mode assignment of the phonon lines in Hg-1234 [34]. The asymmetric line shapes of these phonons result from quantum interference to the electronic counter part. Thus, it is expected that the bare phonon frequency $\Omega_0$ and line-width $\Gamma_0$ are renormalized by the electronic response $\chi(\omega)=R+i\rho$ through the electron-phonon coupling V; $\Omega=\Omega_0+V^2R$ and $\Gamma=\Gamma_0+V^2\rho$, and that the frequency and line-width anomalies below $T_c$ are induced by the $A_{1g}$ electronic response peaking at 580 $cm^{-1}$. To estimate this phonon self-energy effect correctly, one have to analyze the Raman spectrum using the Green's function approach [35-38] or sophisticated Fano approach [39, 40]. In both approaches one introduces an analytical function to describe the background electronic response $\rho$. In the present case, however, the background consists of two components with different resonance behaviors. One exhibits a broad spectrum peaking at about 300 $cm^{-1}$ and the other, which mainly interacts with the relevant phonons, shows a sharp pair-breaking peak at 580 $cm^{-1}$. For the time being, we use a "standard" Fano approach [33, 34, 37]. As was shown in our previous paper [37], for a single component background the "standard" Fano procedure and the Green's function analysis give essentially the same renormalized phonon parameters (frequency, line-width and intensity) both above and below $T_c$, if we assume an appropriate electronic background function $g(\omega)$ in the Fano analysis. In the following analysis $g(\omega)$ is assumed to have a steep slope, reflecting a pair-breaking peak near this phonon.

Figure 5 shows the temperature dependence of the phonon parameters for the 390 $cm^{-1}$ phonon mode at $E_{exc}=2.18eV$. Remarkable softening and broadening occur below $T_c$, owing to development of the $2\Delta(A_{1g})$-peak at 580 $cm^{-1}$. The large amount of softening (30$cm^{-1}$; 8% of the frequency) demonstrates the strong interaction with the electronic system. Similar strong softening and broadening were also observed in Hg-1234 by Hadjiev *et al.* [34]. It may be worth to note that the largest softening of the Ag-phonon in double layer Y-123 is only 3% [41], while it is negligibly small in mono-layer Hg-1201 [42]. The asymmetry parameter q is described as $q=[V(T_p/T_e)+V^2R]/\Gamma$, where $T_p$ and $T_e$ are the phononic and electronic Raman matrix elements. The small $|q|$ above $T_c$ and the large $|q|$ below $T_c$ in Fig. 5 illustrate the asymmetric line shape above $T_c$ and nearly symmetric below $T_c$, respectively. Since the second term contribution is estimated from $\Omega$ and $\Gamma$ as $[(V^2R)_{110K}-(V^2R)_{10K}]/\Gamma \approx 1.3$, the large change in $|q|$ below $T_c$ predominantly originates from the first term.

In Fig. 6, we present the 10K-Raman scattering from the samples of polycrystalline Hg-1234, Hg-1223, Tl-1223 and the single crystal of Y-123 at red (1.92eV) and blue (2.54 eV) laser excitations. For Hg-1234, the strong enhancement of the Ca-phonon peaks, accompanied by the strong $2\Delta(A_{1g})$-peak at $600\,cm^{-1}$, was first observed at $E_{exc}=1.92eV$ by Hadjiev $et\,al.$ [34]. We found that this enhancement is suppressed when $E_{exc}$ increases, namely, that this is a resonance phenomenon. From Fig.6, it is concluded that resonance behavior of the $2\Delta$-peak and the Ca-phonons is commonly observed in HTSC with three or four $CuO_2$-layers in a formula unit, but not in the compounds with a double $CuO_2$-layer such as Y-123 or Bi-2212. Therefore, the superconductivity-induced resonance is closely related to the presence of the inner $CuO_2$-layer sandwiched by the outer $CuO_2$-layers. Moreover, since this phenomenon is observed not only in the optimal doping level (Bi-2223 and Hg-1223) but also in the underdoped Hg-1234 and the overdoped Tl-1223, it turns out that the phenomenon is little sensitive to the average doping level for the multiple layers.

The importance of the inner $CuO_2$-layer and the Ca-phonons in the present resonance phenomenon tells us that the interband Raman scattering plays a major role. The Ca-atoms are located just between the inner- and outer-$CuO_2$ layers and thus $A_{1g}$ Ca-phonons possibly mediate the interband electronic excitations. If the bands originating from the two $CuO_2$-layers were degenerated, the interband part should vanish. Therefore, the present results indicate that the energy bands for the inner and the outer $CuO_2$-layers are slightly different. In this Raman process an electron-like (or hole-like) quasiparticle is created above the gap $\Delta_{inner}$ of the inner $CuO_2$-layer, whereas a hole-like (or electron-like) quasiparticle is above the gap $\Delta_{outer}$ of the outer $CuO_2$-layer. From our results it can be deduced that the gap energy $\Delta_{inner}$ for the inner-layer is almost the same as $\Delta_{outer}$ for the outer-layers.

At the blue excitation the interband Raman scattering is very weak, compared to the intraband channels. It is activated by the resonance. A plausible candidate for the resonance is the one with the optical transition to the upper Hubbard band (~2eV), the oscillator strength of which may survive owing to the underdoped nature of the inner $CuO_2$-plane [43, 44]. Recently, it has been reported from the nuclear magnetic resonance experiments that the doping level of the inner layer is lower than the outer layer in Cu- and Hg-based three (or four)-layer HTSC [5]. The reason why the resonance of the interband Raman process appears only below $T_c$ is currently unclear. The incoherent c-axis conduction in the normal state turns to a coherent transport below $T_c$ owing to a Josephson coupling between the $CuO_2$-layers, which may help the interlayer Raman process. As a possible change in the resonance condition, a superconductivity-induced change in dielectric function in the visible light region has been reported on Bi-2212 by some optical measurements [45]. This might be related to the present observation of diminishing a resonance behavior below $T_c$.

## IV. CONCLUSIONS

In conclusion, we observed resonant Raman scattering below $T_c$ in the $A_{1g}$-spectra of three or four $CuO_2$-layer HTSC. For Bi-2223, the $A_{1g}$ pair-breaking peak shifts from $450 \text{cm}^{-1}$ ($\approx 6k_BT_c$) to $580 \text{ cm}^{-1}$ ($\approx 8k_BT_c$), when the excitation energy changes from 2.54 eV to 2.18 eV. As a result, in resonance, the relative positions of the pair breaking peaks are $2\Delta(A_{1g}) =$ $2\Delta(B_{1g}) > 2\Delta(B_{2g})$, as was theoretically predicted for $d$ – wave multi-layer superconductors but not observed previously. In addition to the changes in the $2\Delta(A_{1g})$ peak position, the intensities of the $2\Delta(A_{1g})$ peak and the Ca-phonons show a remarkable enhancement. We attribute these pronounced spectral changes to the appearance of an interband Raman scattering process that is mediated by the Ca-phonons, resulting from a coupling of the slightly different energy bands for the inner- and outer-$CuO_2$-layers. The resonance at around 2eV excitation energy enhances the unscreened Raman scattering due to multi-sheet effects. This result emphasises recalls us to the importance of resonant Raman studies from both the experimental and theoretical viewpoints.

## ACKNOWLEGMENTS

The authors thank Y. Fudamoto and Yu. Eltsev for experimental assistance and J. Quilty for helpful discussions. This work was supported by the New Energy and Industrial Technology Development Organization (NEDO) as Collaborative Research and Development of Fundamental Technologies for Superconductivity Applications.

**FIGURE CAPTIONS**

Fig.1 $A_{1g}$ Raman spectra of Bi2212 for $E_{exc}$=2.54eV (a) and for $E_{exc}$=2.18eV (b). $A_{1g}$ Raman spectra of Bi2223 for $E_{exc}$=2.54eV (c) and for $E_{exc}$=2.18eV (d). In all cases, the spectra are shown above and below $T_c$ by thin lines and thick lines, respectively. All spectra are corrected for the Bose-Einstein statistical factor.

Fig.2 (a) Temperature dependence of the Raman spectra of Bi2223 for $E_{exc}$=2.18eV. All spectra are corrected for the Bose-Einstein statistical factor. (b) Temperature dependence of the height of the $2\Delta(A_{1g})$-peak at 580cm$^{-1}$ ($I_{580cm-1}$ / $I_{1000cm-1}$) in Bi-2223 spectra for $E_{exc}$=2.54eV and for $E_{exc}$=2.18eV.

Fig.3 (Color) (a) $A_{1g}$ Raman spectra of Bi2223 at T=10K with various excitation lines. Here the continuum intensity at 1000cm$^{-1}$ is assumed to be unity. (b) $A_{1g}$, $B_{1g}$, and $B_{2g}$ spectra with $E_{exc}$ =2.18eV at T=10K.

Fig.4 (a) The height of the $2\Delta(A_{1g})$-peak at 580cm$^{-1}$ ($I_{580cm-1}$ / $I_{1000cm-1}$) in Bi-2223 spectra, compared with the height of the $2\Delta(A_{1g})$ peak at 390 cm$^{-1}$ ($I_{390cm-1}$ / $I_{1000cm-1}$) in Bi-2212 spectra. (b) The resonance Raman profiles of the renormalized phonon intensities $I_p$ for the 260 and 390 cm$^{-1}$ phonon lines in Bi-2223 spectra. T=10K. The lines are guides to the eye.

Fig.5 Temperature dependencies of the phonon parameters of the 390 cm$^{-1}$ phonon for the Bi-2223 single crystals at $E_{exc}$ =2.18eV: (a) renormalized frequency $\Omega$, (b) renormalized linewidth $2\Gamma$, (c) the Fano parameter -q. The solid lines display the expected T-dependencies for the uncoupled frequency and linewidth [37].

Fig.6 The 10K-Raman spectra of the polycrystalline Hg-1234 (a), Hg-1223 (b), Tl-1223 (c), and the Y-123 single crystal (d) obtained at $E_{exc}$ =1.92eV (solid curves) and 2.54eV (dashed curves). In (d) panel, the spectrum has $A_{1g}$ symmetry, while the spectra in (a, b, c) panels are polarized ones. Intensities of the data for 2.54eV in (d) are multiplied by 3.

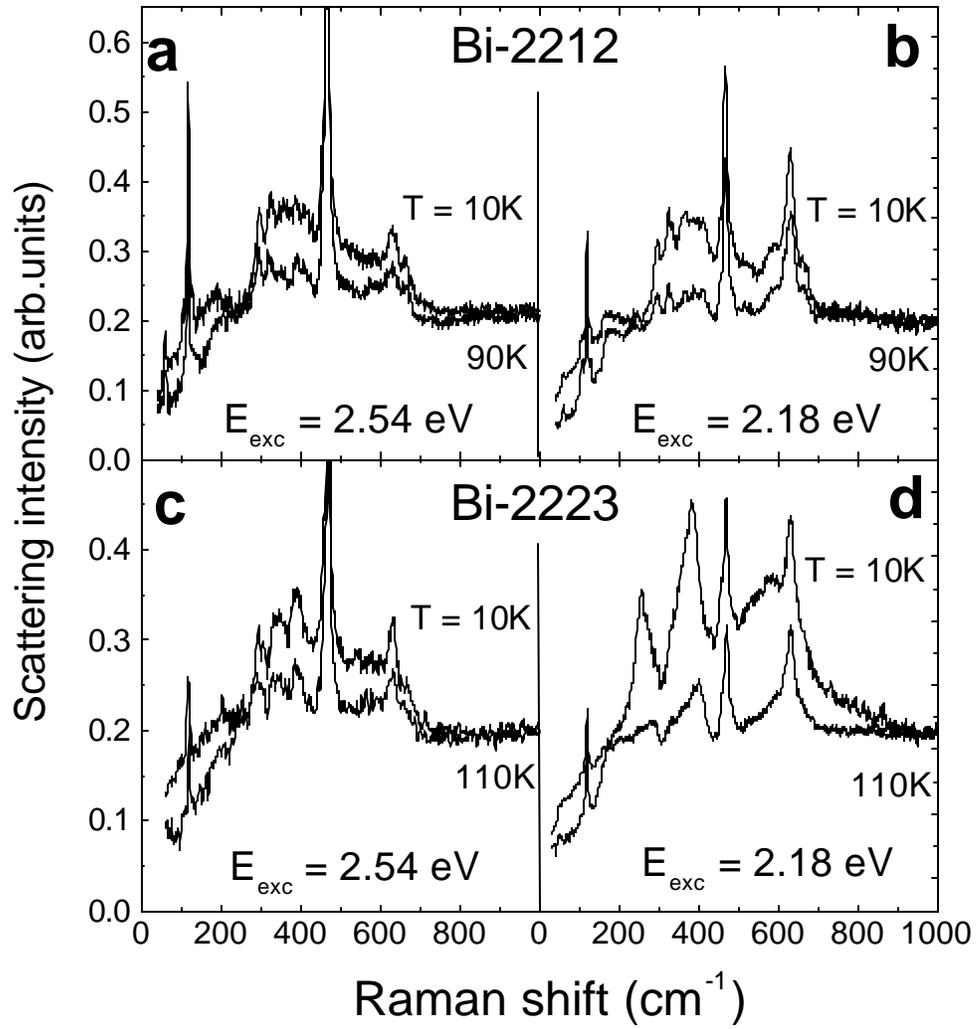

Fig.1 Limonov e.a.

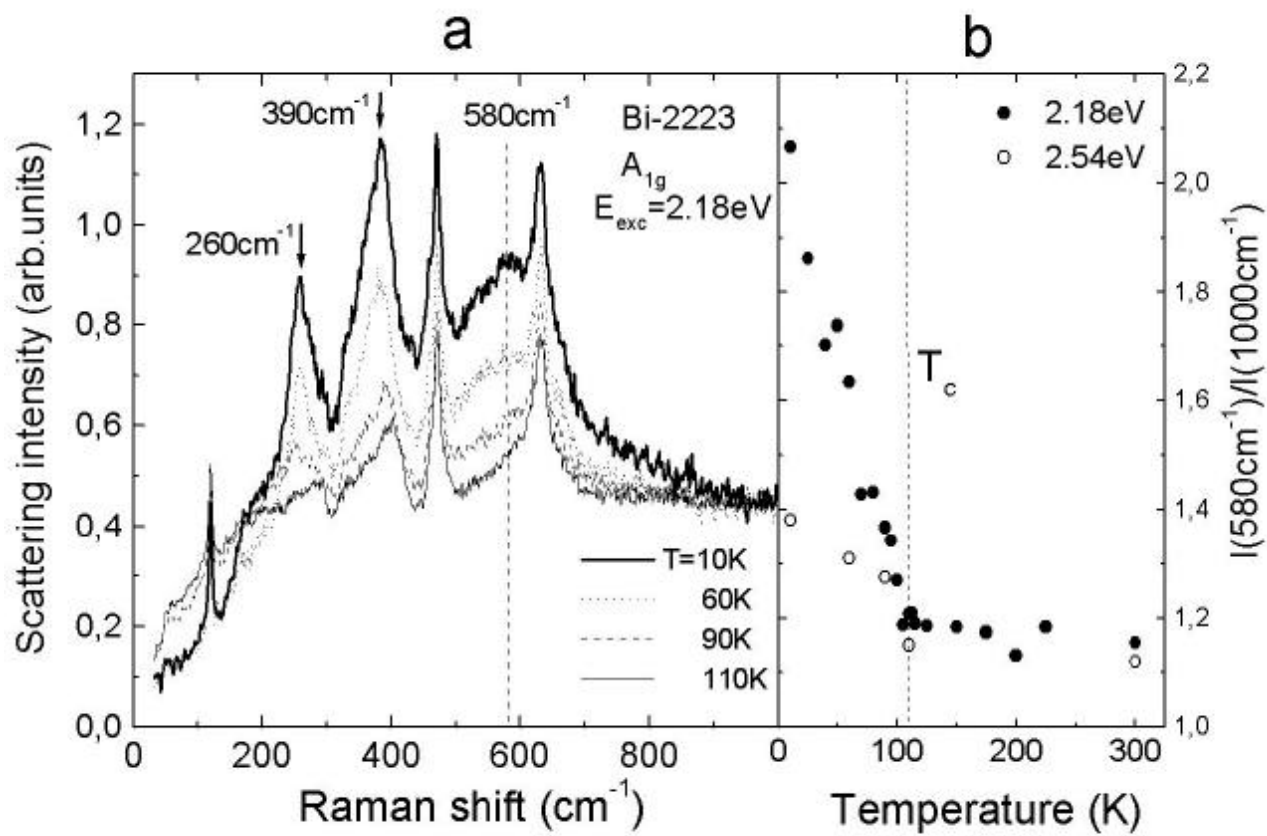

Fig.2 Limonov e.a.

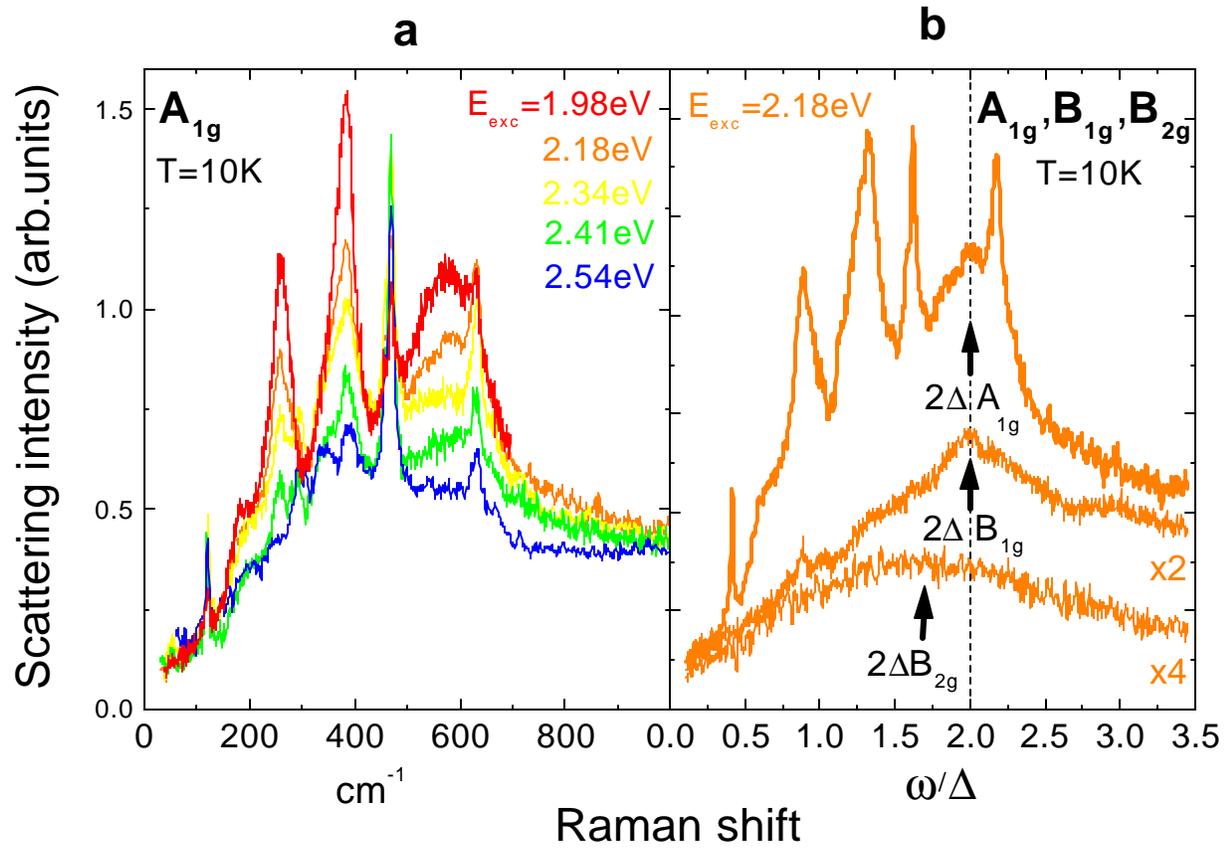

Fig.3 Limonov e.a.

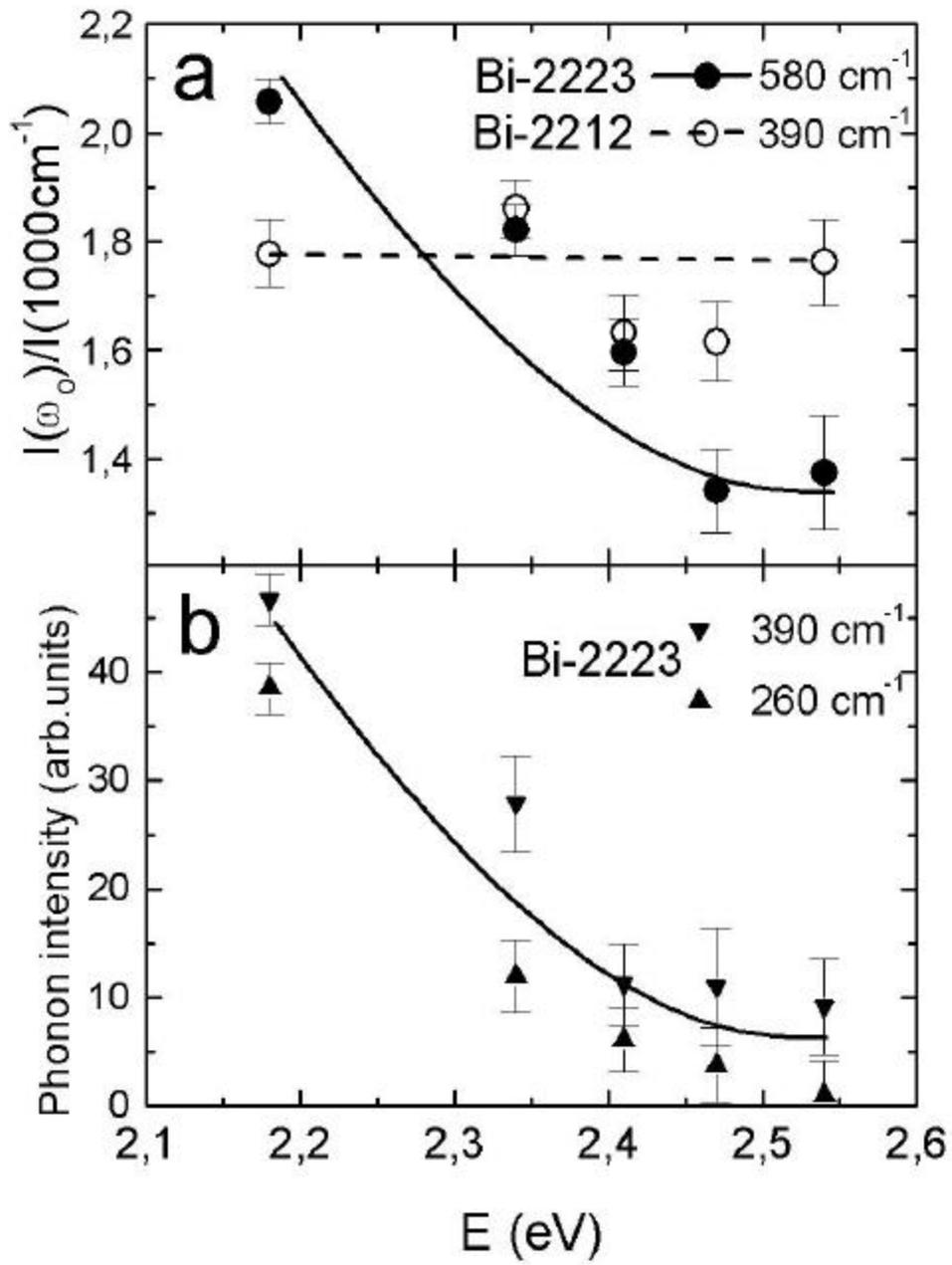

Fig.4 Limonov e.a.

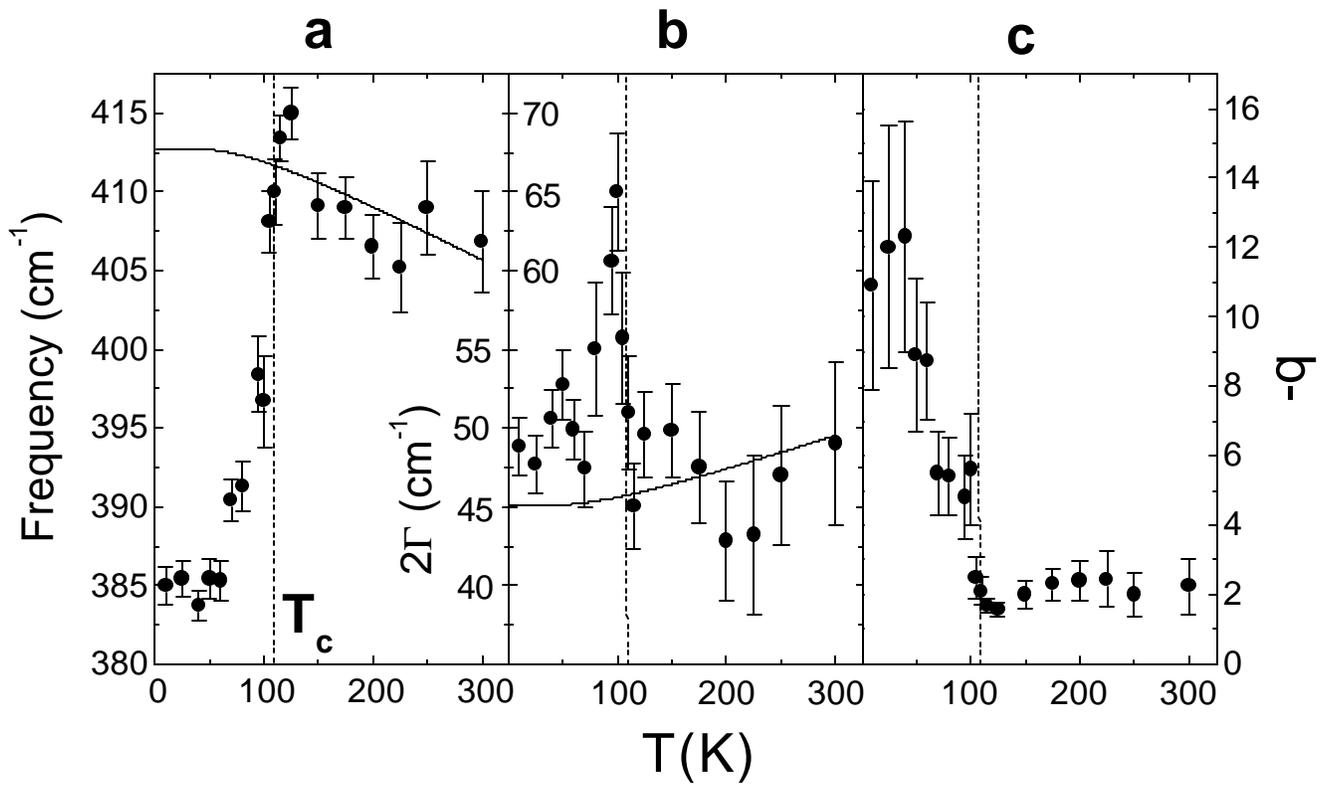

Fig. 5 Limonov e.a.

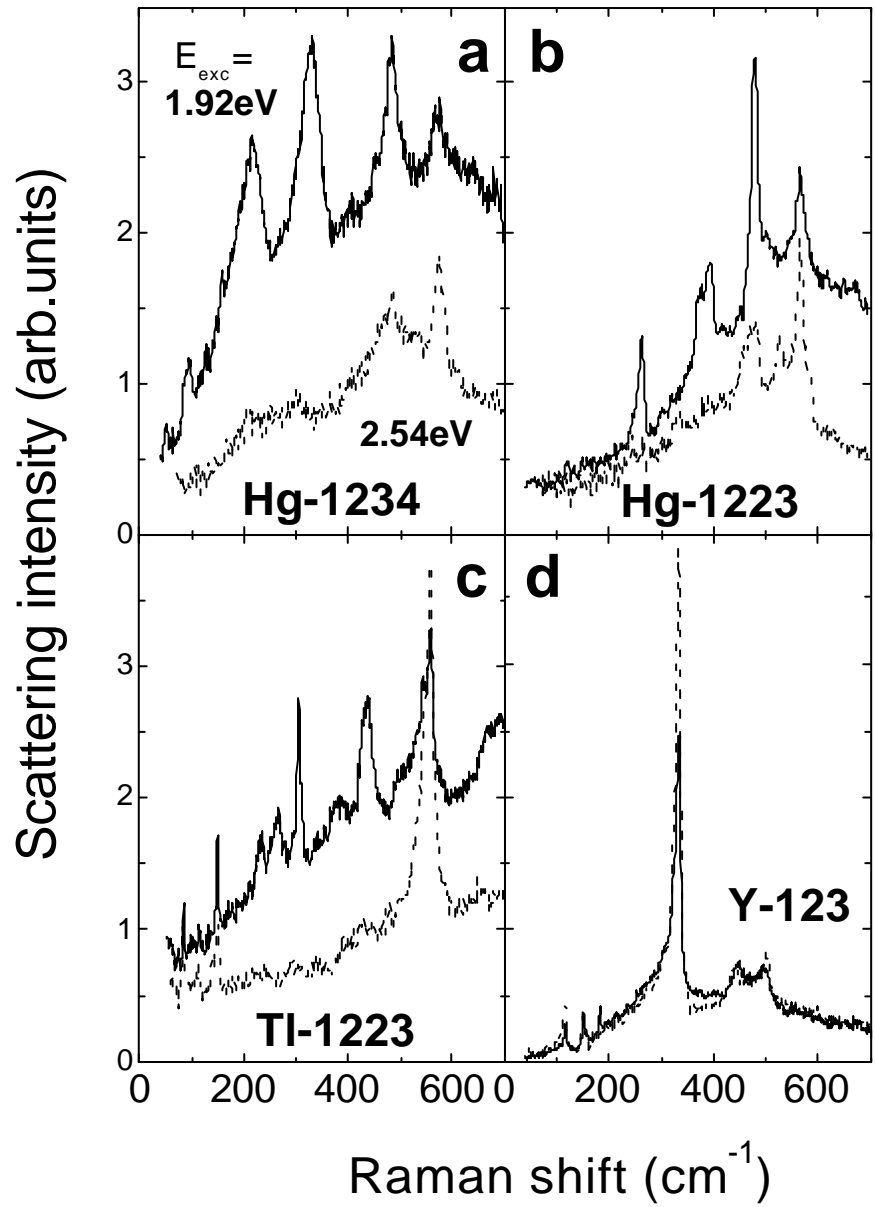

Fig.6 Limonov e.a.